# Are We Testing or Being Tested? Exploring the Practical Applications of Large Language Models in Software Testing

Robson de Souza Santos
UNINASSAU
Triunfo, PE, Brazil
robsonrtss@gmail.com

Italo Santos
Northern Arizona University
Flagstaff, AZ, US
ids37@nau.edu

Cleyton Magalhaes
CESAR School
Recife, PE, Brazil
cvcm@cesar.school

Ronnie de Souza Santos
University of Calgary
Calgary, AB, Canada
ronnie.desouzasantos@ucalgary.ca

*Abstract*—A Large Language Model (LLM) represents a cutting-edge artificial intelligence model that generates coherent content, including grammatically precise sentences, human-like paragraphs, and syntactically accurate code snippets. LLMs can play a pivotal role in software development, including software testing. LLMs go beyond traditional roles such as requirement analysis and documentation and can support test case generation, making them valuable tools that significantly enhance testing practices within the field. Hence, we explore the practical application of LLMs in software testing within an industrial setting, focusing on their current use by professional testers. In this context, rather than relying on existing data, we conducted a cross-sectional survey and collected data within real working contexts—specifically, engaging with practitioners in industrial settings. We applied quantitative and qualitative techniques to analyze and synthesize our collected data. Our findings demonstrate that LLMs effectively enhance testing documents and significantly assist testing professionals in programming tasks like debugging and test case automation. LLMs can support individuals engaged in manual testing who need to code. However, it is crucial to emphasize that, at this early stage, software testing professionals should use LLMs with caution while well-defined methods and guidelines are being built for the secure adoption of these tools.

## I. INTRODUCTION

A Large Language Model (LLM) is an advanced artificial intelligence model extensively trained on diverse data sources, including books, code, articles, and websites. LLM models, such as GPT-3, BERT, T5, and XLNet, explore inherent patterns and relationships within the language they are trained on, allowing them to produce coherent content, including grammatically accurate sentences, human-like paragraphs, and syntactically correct code snippets [1]. LLMs represent a substantial advancement in natural language processing and machine learning. They recently gained popularity across various disciplines due to their proficiency in understanding and generating extensive and contextually relevant language [2].

In fields like Healthcare, LLMs can extract key terms from medical documents, recover missing patient data, redact sensitive information, and enhance patient communication through advanced text analysis [3]. Within education, LLMs seamlessly enhance several processes, encompassing automated grading, class content creation, personalized learning materials, and accessibility tools, fostering a dynamic learning experience [4]. In finances, LLMs can be applied for risk assessment and fraud detection through sentiment analysis [5]. Indeed, software engineering, at the forefront of innovation, strategically employs LLMs to refine practices and processes, resulting in substantial advancements within the field [1].

In software engineering, LLMs can support several activities within the software development process. Among the tasks a software team performs, they can refine software specifications, generate code, support unit testing, and contribute to creating comprehensive software documentation [1], [6]. Such versatile application supports project management and planning strategies and promotes an efficient knowledge-building approach to the software development process.

Considering specifically software testing and the activities around software quality, LLMs assume a significant role beyond mere requirement analysis and documentation [1]. They extend their impact beyond unit testing, emerging as powerful tools for test case generation [7], hence offering robust support in optimizing testing practices within the field. A recent study explored the literature on this topic and demonstrated that the applicability of LLMs in software testing spans a range of tasks, including test case preparation, bug reporting, and bug fixing [8].

In this sense, considering the potential applicability of LLMs in software testing and quality assurance practices, in this study, we investigate the use of LLMs by professional testing professionals within the industrial context, guided by the following research question:

*RQ: How are software professionals using LLMs in their testing activities?*

From this introduction, our study is organized as follows. Section II presents a theoretical background on software testing and related work. In Section III, we describe our method. In Section IV, we present our results, which are discussed in Section V. Finally, Section VI summarizes the contributions of this study.

## II. BACKGROUND

In this section, we present concepts and definitions of software testing and discuss the findings published in [8] that explore using LLMs in software testing through the lenses of previous studies.

### A. Software Testing

Software testing is an essential aspect of software quality assurance, involving a systematic process of planning, designing, and executing testing activities. Testing a system involves systematically evaluating its behaviors to identify bugs, errors, or inconsistencies in an interactive process that helps to ensure that the software meets quality standards and functions as intended before reaching end-users. Hence, software testing can save costs in the development process by preventing defects early, reducing rework, and enhancing user satisfaction [9]–[13].

The software testing life cycle is a dynamic set of activities designed to verify and validate a software product. This process is intrinsically incorporated into the general software development life cycle, and its adaptability allows professionals to integrate it with diverse methodologies like agile or waterfall. Regardless of the chosen approach, certain common software testing life cycle activities persist across methods, processes, and team configurations [14]–[16]. In summary, the software testing life cycle typically includes [17]:

- *Requirements Analysis*: Concentrates on a comprehensive understanding and analysis of testing requirements derived from software specifications, establishing a robust foundation for the testing process.

- *Test Planning*: Focuses on crafting an extensive test plan that details the testing strategy, scope, objectives, resources, and schedule. This process creates a clear roadmap to guide the testing process effectively.

- *Test Case Design*: Centered on creating detailed test cases spanning various scenarios, conditions, and inputs to thoroughly assess the software's functionality.

- *Test Execution*: Focused on implementing test cases, using manual or automated testing tools to identify defects, validate software functionality, and ensure compliance with the specified requirements.

- *Regression Testing*: Centered on retesting previously validated functionalities to verify that recent changes or fixes have not unintentionally introduced new defects or impacted existing features.

- *Bug Reporting*: Focused on the detailed documentation and reporting of identified defects, offering developers the essential information to address and fix issues effectively.

Software professionals apply various techniques and tools in implementing the aforementioned activities. This includes employing test design methods to broaden coverage, utilizing automation tools to enhance test execution, integrating performance testing and continuous integration tools, and fostering collaboration through diverse platforms to improve communication and ensure a thorough understanding of requirements and bugs [16], [18], [19].

In contemporary software development trends, software professionals, including those who work specifically with software testing, are embracing LLMs as integral tools to support them in their work. This resonates with a prevailing sentiment within the software industry, recognizing language models' substantial impact and advantages to various facets of software development [1], [6], [8].

### B. Related Work: Software Testing with Large Language Model [8]

A recent research analyzed 52 relevant studies investigating the use of LLMs in software testing. Spanning the years 2020 to 2023, the number of studies has shown a nearly exponential increase. In 2020 and 2021, there were merely 1 and 3 papers, which increased to 13 papers in 2022, and in the first half of 2023, have already achieved an impressive total of 33 publications. This upward trend suggests growing popularity and increasing attention to this evolving field.

The review demonstrated that LLMs can play an essential role in the initial and late stages of the software testing lifecycle. In mid-phase tasks, such as test case design, they facilitate the creation of test oracles, the implementation of unit tests, and the generation of test data. Transitioning to later stages, including bug reports and regression testing, LLMs effectively support fault analysis, debugging, and repair.

Additionally, the study indicates that Codex is the most widely utilized LLM tool among software testers, as it is trained on a diverse code corpus, including languages like JavaScript, Python, C/C++, and Java, serving as an AI pair programmer that generates complete code snippets from natural language prompts. ChatGPT emerges as the second most used LLM, employed in various software testing tasks, particularly for generating human-like text responses for data inputs. Lastly, the third-ranked LLM, CodeT5, is utilized for pre-training and fine-tuning with domain-specific data, particularly in program repair and unit testing.

## III. METHOD

In this research, we opted for a different approach from [8] to investigate how testing professionals use LLMs in their activities. Rather than relying on existing data, we collected our data within real working contexts—specifically, engaging with practitioners in industrial settings. With this strategy, we aimed to offer a distinctive and complementary perspective to enhance the understanding of LLM usage in the field.

In this sense, we conducted a cross-sectional survey [20] following well-established guidelines in software engineering [21] to explore LLMs usage in software testing. Initially, we devised a questionnaire (III-A) based on the existing literature on using LLMs in software testing, particularly based on [8]. Following this, we used Prolific[1] to spread the questionnaire among software testing professionals around the world (III-B). Then, we applied quantitative and qualitative techniques to analyze and synthesize our collected data (III-C). Below, we describe our method in detail.

### A. Questionnaire

We created an anonymous survey. The survey consisted of four (4) main sections: (i) PART I - LLMs in Software Testing; (ii) PART II - Using LLMs in Specific Tasks; (iii) PART III - Potential of LLM in Software Testing; and (iv) Part IV - Demographics. We designed the questions to understand how testing professionals apply LLMs during work activities. We used open and multiple-choice questions. Our survey instrument is provided in the replication package[2]. Below, we briefly explain the survey sections:

- *Part I - LLMs in Software Testing.* Participants were asked to define the main activities related to software testing and software quality in their work, specify any LLM tools, websites, applications, or algorithms they used to support their work, and briefly describe how they incorporated LLMs into their tasks.

- *Part II - Using LLMs in Specific Tasks.* In this section, we inquired whether participants utilized LLMs for executing specific tasks such as requirements analysis, test plan creation, test design and reviews, test case creation, test execution, bug reporting, bug fixing, regression testing, or software release.

- *Part III - Potential of LLMs in Software Testing.* We asked participants to share their thoughts about how they think LLM tools can be used to understand, generate, and manipulate textual content. Then, we asked what activities within software testing they believe LLMs would be more helpful.

- *Part IV - Demographics.* Participants were asked about gender, the country where they live, years of experience with software testing/software quality, the kind of projects they are working and if they have experience with agile development.

Besides these four main sections, the survey starts with *Pre-Screening Validation* section questions to validate if the participants recruited are part of the sample we wanted to participate in our study. After this section, we have a consent page explaining to participants the purpose of the research, the confidentiality rules, the time needed to complete the survey, and the researcher's contact. Furthermore, an attention check question was also included to ensure that participants read instructions carefully, as recommended by Danilova et al. [22]. An attention check question [23] appeared randomly during the survey, and it is essential to filter out careless respondents. For the attention check question, we use: When asked if participants know about software testing, answer "Tester". The questions and answer options were randomized to mitigate response fatigue and response order effects [24].

### B. Data Collection

Recruiting participants for empirical software engineering investigations presents a common challenge, as noted by [25]. To address this, we chose to widely distribute our survey while adhering to ethical considerations outlined by OSS communities [26]. We refrained from scraping email addresses from software repositories, a practice that may infringe upon platform terms of service and regional data protection laws. Our focus was on increasing the response rate and enhancing the diversity of our sample. To achieve this, we leveraged the Prolific platform to promote our survey, and the forthcoming section will elucidate the procedures employed for participant recruitment on this platform.

Prolific[3] is a crowdsourcing platform, and you can channel your survey through it. By January 2022, Prolific had more than 150,000 active users [25]. We administered our surveys on Google Forms [4]. We conducted our survey using Prolific, considering the steps followed by some works that successfully recruited participants using this platform [22], [25]. The prolific first selection process starts when participants are self-selected by providing several demographics. Prolific provides some pre-screening questions to help to narrow down the relevant populations. As our target population is testing professionals, we selected the following filters: (i) Knowledge of software development techniques, filtered for functional testing, A/B testing, and unit testing; (ii) Programming skills; and (iii) Technology use at work, filtered for 4 or 6 times a week. By September 2023, according to Prolific information, we had 554 matching participants who had been active on the platform for the past 90 days.

Unfortunately, we can not be sure those participants are part of our target population. They could be lying about the information provided to the platform to win money by participating in the surveys. Thus, we included more screening questions to improve the quality of our study. To do this, we have a pre-screening validation section with the same questions asked by the prolific platform filters selected and detailed previously. Besides those questions, we also included the recommended questions for Danilova et al. [22] that developed a list of questions to help filter students with programming skills. We adapted some questions from her work and included a question to validate if the participants knew basic concepts about software testing. After respondents answer the pre-screening validation section, they will have access to the complete survey if they pass those questions. If not, they were redirected to the end of the study. The survey was available between September 1 and October 15, 2023.

---

[1]www.prolific.com
[2]https://figshare.com/s/f4209b4e7ebd1ec10ce0
[3]www.prolific.com
[4]https://www.google.com/forms/

We received 212 non-blank answers, and after filtering the data (as detailed in the following subsection), we ended up with 83 valid responses.

### C. Data Analysis

We started data analysis by carefully reviewing and filtering our data to consider only valid responses. We dropped answers that failed the attention check question (22 cases). We then dropped incomplete questionnaires (105 removed). We manually inspected the open-text questions for senseless and inappropriate answers (2 removed). Moreover, we filtered our data for potential duplicate participation, even though the Google form has mechanisms to prevent multiple responses from the same participant. After applying all the filters, we ended up with 83 valid responses. Once we completed the filtering process, we employed quantitative and qualitative data analysis techniques to summarize and interpret the collected data. For quantitative data, we used descriptive statistics [27], and for quantitative data, we used thematic analysis [28].

Initially, descriptive statistics [27] were employed to describe the key characteristics of our sample quantitatively. Using statistical functions, including means, proportions, totals, and ratios, we organized participants' responses into subgroups, gaining insights into the data. Interactive visualizations, created using Tableau, facilitated the analysis of data patterns such as averages and distributions. Aggregations were employed to explore the experiences of various groups of participants in the sample, considering factors like roles and backgrounds.

Following this, we applied the thematic analysis [28] to explore the detailed responses from participants to open-ended questions. This method facilitated extracting and identifying recurring themes and patterns within the qualitative data, contributing to a comprehensive understanding of participants' experiences and perspectives. Thematic analysis was crucial in revealing nuanced insights and enhancing the qualitative dimensions of our study.

### D. Ethics

Following ethical principles, no personal information about the participants was collected in this study (e.g., name, e-mail, or employer) to maintain participants' anonymity. Before starting the questionnaire, participants were informed about the study's goals and had to agree to use their answers for scientific purposes.

## IV. FINDINGS

We obtained 83 completed questionnaires. Our participant group is diverse, with experienced individuals from three continents involved in various system domains. Engaging in activities such as functional testing, unit testing, A/B testing, and quality assurance, many participants have experience in agile environments. For a detailed understanding of our sample, Section IV-A provides comprehensive insights into the characteristics of these seasoned individuals.

### A. Demographics

Our diverse sample includes participants from 21 countries, with a notable representation of professionals from the United Kingdom. Our sample exhibits a balanced gender distribution, comprising 35% women. A significant portion (over 67%) of our participants have experience in functional testing, while close to 63% have engaged in unit testing. The majority of participants (38%) specialize in web application development, and more than 70% work in agile environments. Finally, 48% of participants indicated using LLMs to support their testing activities. Refer to Table I for a comprehensive quantitative breakdown of our sample distribution.

TABLE I
DEMOGRAPHICS

| Participants Profile | | |
|---|---|---|
| Gender | Male | 65% |
| | Female | 35% |
| Location | United Kingdom | 25% |
| | South Africa | 20% |
| | United States | 12% |
| | Portugal | 5% |
| | Netherlands | 5% |
| | Germany | 5% |
| | France | 4% |
| | Canada | 4% |
| | Spain | 2% |
| | Poland | 2% |
| | Mexico | 2% |
| | Italy | 2% |
| | Greece | 2% |
| | Scotland | 1% |
| | Nigeria | 1% |
| | Israel | 1% |
| | Hungary | 1% |
| | Czech Republic | 1% |
| | Belgium | 1% |
| | Australia | 1% |
| Experience | 0-2 Years | 40% |
| | 2-5 Years | 36% |
| | 5-10 Years | 17% |
| | 10+ Years | 7% |
| Agile Software Development | Yes | 70% |
| | No | 30% |
| Software Project Domain | Web | 38% |
| | Desktop | 34% |
| | Mobile | 23% |
| | Game | 5% |
| LLMs Usage | No | 52% |
| | Yes | 48% |

### B. Using LLMs in Software Testing

Similar to [8], we queried participants about their usage of LLMs in various activities within the software testing life cycle. As indicated in our demographics (Table I), most participants had not incorporated LLMs into their work until data collection. Nevertheless, among those who did, our analysis revealed that software testing professionals apply LLMs across diverse tasks in multiple stages of the testing process. Applying the classification strategy used by [8], our findings indicate that 40% of professionals utilize LLMs during the initial phases of software testing—encompassing requirements analysis, test design, and test plan creation. Moreover, 28% of professionals in the sample leverage LLMs during intermediate stages, including test case preparation and test execution. Lastly, 32% of participants predominantly use LLMs in post-testing activities, providing support for bug fixing, regression testing, and software release. Figure 1 presents our results using a format similar to the one introduced in [8], providing a clear visual representation of our findings.

Furthermore, exploring qualitative data from open-ended questions enriched our understanding of how software testing professionals integrate LLMs into their practices. This analysis provided a nuanced perspective, offering comprehensive insights into the versatile applications of LLMs in the dynamic landscape of software testing.

For instance, regarding *requirements analysis*, using LLMs seems to be focused on using natural language processing to conduct text analysis in the collected requirements, as commented by P063: *"they can be used for natural language processing tasks, aiding in the analysis of user requirements"*. P037 adds to this discussion by stating that *"these models can analyze software requirements, extract key information"*.

For activities involving *test planning*, software testers seem to be using the power of LLMs to generate texts that can be incorporated into the test plan documentation. As an example, P045 commented that *"LLMs can provide standardized test plan templates or frameworks, ensuring that important sections like objectives, scope, resources, and timelines are included in the plan"*.

In the context of *test design and review*, professionals leverage LLMs to aid them in defining test scenarios and in programming activities, including enhancing test coverage by identifying edge cases producing data input to the tests and coding. P073 reported: *"using the generation of test cases and test data, LLMs can help ensure thorough test coverage"*. P053 experienced LLMs for coding activities: *"I have used an LLM to help generate code for some functions that I was stuck on"*.

Moreover, the professionals reported that LLMs are useful in activities associated with *test case preparation*. Within our sample, we identified strong evidence for using LLMs in testing automation, such as writing code and scripts for automating test cases. For example, P009 commented: *"basically just use Github Copilot to help me spit out tests faster. I use ChatGPT to do the same."*. P077 supplemented the evidence about these findings by stating that: *"LLMs are used to generate well organized and logically structures code"*.

Considering *test execution*, insights from our participants highlight the broad applicability of LLMs and indicate their versatility to assist them in test execution across diverse system domains and user characteristics. This understanding is firmly grounded in the specific examples shared by participants, which offer evidence of how LLMs effectively support testing execution in diverse testing scenarios. P016 provided an example: *"I automate the recruitment process from sourcing candidates to screening resumes"*. P035 also commented on the execution of test cases with LLMs in a particular context: *"We use generative text prediction tools as part of our customer service help desk processes to interact with our customers online"*.

Regarding *bug fix and test regression*, our participants emphasized the role of LLMs in the code debugging process, which includes activities such as locating bugs, making necessary fixes, and handling errors by leveraging suggestions to improve the code. Moreover, our findings indicate that LLMs are used to identify bugs introduced into the system after modifications (i.e., regressions). P036 said: *"I've used GPT-3 based tools such as ChatGPT and GitHub Co-Pilot to help with errors, write boiler code, and give suggestions for my problems"*. P074 indicates using the tools for several debugging and code fixing tasks: *"use a lot for boilerplate code, fixing small bugs, and simple queries regarding syntax for different programming languages"*. Additionally, P10 commented on test regression: *"I used it mostly to evaluate quality and catch regressions"*.

Finally, although two participants in our sample mentioned employing LLMs in software release tasks, their responses needed more evidence with concrete examples of how they apply these tools in their work. Yet, given that many professionals highlighted using LLMs for coding support and document generation, we hypothesize their potential use in generating user guidelines and automating the release process, including tasks involving multiple code branches. Further investigations are necessary to gather concrete evidence and validate these hypotheses.

### C. Perspectives

As previously mentioned, despite working in testing, more than 43% of our sample has not integrated LLMs into their activities. However, discussions on this topic are gaining momentum. Considering that all professionals in our sample, both LLM users and non-users, may hold perspectives on how these tools could support their work in the future, we inquired about their anticipated use of LLMs in testing. Specifically, we asked participants how they envision incorporating LLMs into their testing activities in the near future and which activity in the software testing life cycle they believe LLMs will be most helpful.

Figure 2 presents the distribution of participants' responses, distinguishing those not currently using Language Model Libraries (LLMs) in red and current users in blue. Most testing professionals who do not use LLMs recognize their value in test case generation, a perspective that current users share. Notably, there is a notable increase in professionals who are current users of LLMs, anticipating their application in test case creation. Among non-users, the second most anticipated use of LLMs is in bug report activities, while current users expect to continue leveraging them for bug fix support. Additionally, a substantial number from both groups foresee using LLMs for test design. In summary, our sample envisions the potential benefit of LLMs across all activities in the software testing life cycle in the near future.

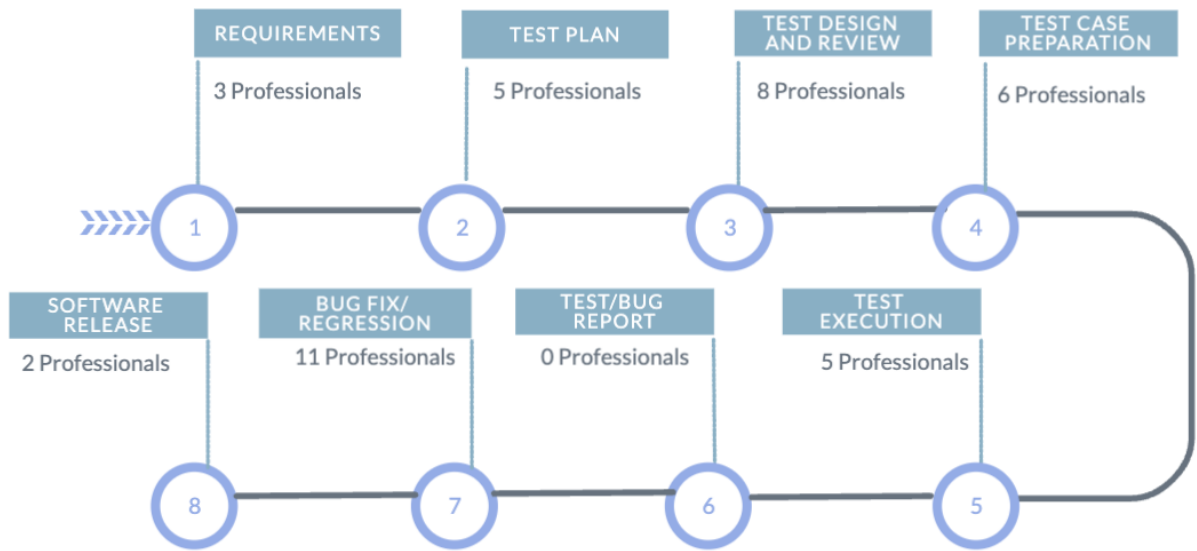

Fig. 1. Uses of LLMs in the Software Testing Life Cycle

## V. Discussion

In this section, we compare our findings and those documented in the review conducted by [8]. Following this, we explore the implications of our results for both research and industrial practices. We conclude our discussions by addressing the limitations of our study and considering potential threats to validity.

### A. Enfolding the Literature

The comprehensive review of 52 studies on LLMs in software testing conducted by [8] lays a strong groundwork. A comparative analysis of our results with theirs promises to offer a comprehensive panorama of how LLMs are employed in this activity process within the software development process.

Broadly, our findings reveal that practitioners in software testing are utilizing LLMs across a more extensive range of activities than previously documented in the literature. It not only corroborates existing results, affirming LLMs' support in test case preparation, bug reporting, bug fixing, and regression testing, but also contributes to the field by providing evidence of their application in requirement analysis, test planning, test design, and test execution—thus enriching our understanding of LLMs' practical implications in software testing.

A key finding in our study is the role of LLMs in supporting testing professionals in their test automation activities. As previously addressed in the literature, coding proficiency, especially for test automation, is increasingly essential for testing professionals in the software industry [29]. Hence, it is noteworthy to acknowledge the role of LLMs in debugging activities and assisting professionals in writing testing code effectively.

In contrast to [8], our study did not delve into the potential challenges encountered by practitioners. However, given the widespread enthusiasm among professionals for using LLMs in software testing, it is foreseeable that challenges may arise as the effective integration of different LLM tools for diverse testing activities might pose difficulties. The lack of a thorough understanding of how these tools impact the work in software engineering could become a substantial problem, especially considering recent issues reported with LLMs, as highlighted in [30].

### B. Implications

Our findings offer valuable insights for practitioners and researchers, enriching our comprehension of using LLMs for software testing.

For academic research, we not only expanded the existing body of knowledge in the literature [8] but also identified opportunities for further investigations. We recommend researchers focus on exploring the effective use of LLMs in aspects of software testing that have received limited attention, including requirements analysis, test reporting, and software release. We emphasize the abundance of information in the grey literature, such as forums and social media, guiding practitioners in LLM usage. However, it is crucial to approach these sources cautiously and subject them to rigorous scientific methodologies.

In terms of industrial practice, we have provided examples illustrating practitioners' current global utilization of LLMs. Our findings can be an initial basis for discussions across different software contexts. Specifically, we highlight the usage of LLMs in supporting testing activities involving programming, such as debugging and automating test cases. Software testing practitioners can consider LLMs valuable supporting tools to enhance their work. However, in light of recent issues reported in the literature [30], we advise practitioners to exercise

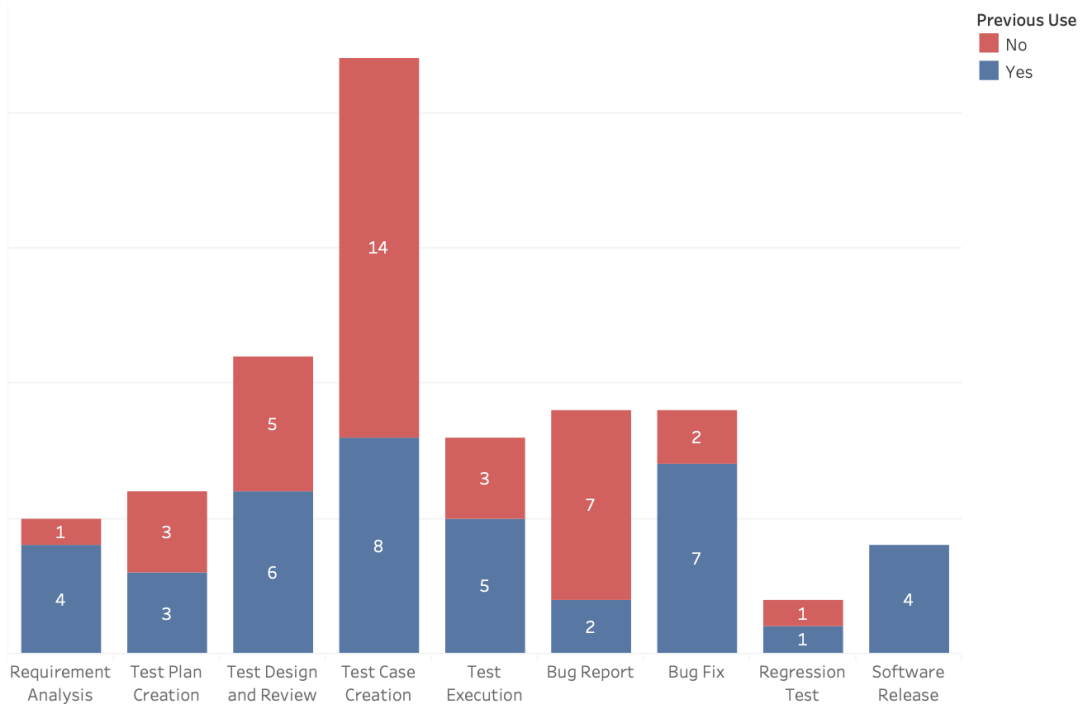

Fig. 2. Perspectives of the Use of LLMs in the Software Testing

caution, mainly when dealing with tests that involve sensitive information.

*C. Limitations*

At this point, we must acknowledge certain limitations in our study, particularly concerning potential threats to validity. Initially, our study was designed with a primary focus on software professionals specializing in testing activities, encompassing roles such as software testers, test engineers, test analysts, and quality assurance specialists. However, we recognize a limitation in the filters used in Prolific filters that led to the inclusion in our sample of developers who also perform testing-related tasks, such as unit testing. While this poses a limitation, we recognize the role developers play in quality activities today. Consequently, incorporating their perspectives has enriched the applicability of our findings. Secondly, as commonly expected in qualitative studies, it is essential to highlight that our findings are not generalizable to a broad population in a positivist manner. Nonetheless, we assert that our findings can be re-analyzed and transferred to various contexts. Furthermore, our discussions can catalyze meaningful knowledge exchange, providing valuable insights to practitioners.

*D. Future Works*

To the best of our knowledge, this study stands as the first empirical research delving into the experiences of practitioners in the industrial context regarding the use of LLMs in software testing. Our findings highlight a consistent and growing utilization of LLMs in software testing activities, as perceived by the professionals in our study. Building on these results, we recognize the need for the next steps in this research that involve developing guidelines that effectively support practitioners in selecting and applying these tools in their activities, which includes aspects related to validating efficacy and ensuring appropriate usage to mitigate issues inherent to LLMs that could compromise software quality.

## VI. CONCLUSION

With the rising popularity of LLMs, it is only natural for these tools to gain attention in the software industry, including in providing support to professionals engaged in various software development activities. Specifically, in the software testing life cycle, practitioners indicate that LLMs can be valuable across all software verification and validation stages. These include initial activities like requirements analysis and intermediate processes such as test execution and extend to the final stages, supporting bug fixing and software release. This research identifies evidence and insights from testing professionals using LLMs in various testing activities.

In summary, LLMs can effectively support the improvement of documents used in the testing process, and more importantly, these tools help testing professionals in programming tasks, such as debugging and test case automation, which can be challenging for those involved in manual testing, for instance. However, we emphasize that, at this initial stage, software testing professionals should approach LLMs with caution until well-defined methods and guidelines can securely guide the adoption of these tools.